\documentclass[prl,twocolumn,superscriptaddress]{revtex4}

\usepackage{epsfig}

\newcommand{\ran}{\rangle}
\newcommand{\kth}{k^{\rm th}}
\newcommand{\tlambda}{\tilde{\lambda}}
\newcommand{\tGamma}{\tilde{\Gamma}}

\begin{document}

\title{Classical simulation of limited-width cluster-state quantum computation}

\author{Nadav Yoran}\email{N.Yoran@bristol.ac.uk} \author{Anthony J. Short}
\affiliation{H.H.Wills Physics Laboratory, University of Bristol,
Tyndall Avenue, Bristol BS8 1TL, UK}


\begin{abstract}
We present a classical protocol, using the matrix product state
representation, to simulate cluster-state quantum computation at a
cost polynomial in the number of qubits in the cluster and
exponential in $d$ -- the width of the cluster. We use this result
to show that any log-depth quantum computation in the gate array
model, with gates linking only nearby qubits, can be simulated
efficiently on a classical computer.
\end{abstract}

\maketitle

The model of cluster-state quantum computation proposed by
Raussendorf and Briegel~\cite{rb} has attracted much interest in
recent years. This model demonstrates some remarkable features,
most notably the fact that once a particular multi-qubit state
(the cluster-state) has been prepared, the whole processing of the
data is implemented by measurements of individual qubits and
feed-forward (that is, the measurement carried out on one qubit
may depend on the outcomes of previous measurements). In general
the cluster-state is thought of as a two-dimensional grid of
qubits, entangled by applying CPHASE operations between
neighbouring pairs. Due to this entangled structure, any quantum
circuit can be simulated by cluster state computation~\cite{rb}
with a polynomial resource overhead (in terms of the number of
qubits and elementary operations). Yet, the computational power of
a cluster does not depend only on the amount of resources that
were invested in its construction, but also on its geometry. It
was shown by Nielsen that any computation implemented with a
linear cluster (a single chain of qubits) can be simulated
efficiently on a classical computer~\cite{niel}.

In this paper we analyse the computational power of limited-width
cluster-states. We show that any computation on a cluster-state in
the shape of a rectangular grid can be simulated on a classical
computer at a cost which is quadratic in the number of qubits and
polynomial in $2^{d}$, where $d$ is the width of the cluster.
Therefore, if we limit the width so it scales like the $\log$ of
the number of logical qubits, then any computation on that cluster
can be simulated efficiently. We will generalize our proof to
include also clusters in which the connections between qubits are
not necessarily between nearest neighbours, but are bounded by
some constant length.

Since any quantum computation can be implemented by cluster-state
computation our results imply that any quantum computation in the
gate array model with depth that scales like $\log$ of the number
of qubits, and where the range of the interactions is bounded by a
constant (that is two-qubit gates are applied only to qubits which
are not too far apart) can be efficiently simulated on a classical
computer. Similar results concerning the computational power of
limited-depth quantum gate arrays have been reported recently
\cite{ms} following from a completely different approach.

Our approach relies on the representation of the cluster-state as
a matrix product state (MPS). We shall follow here the
construction of Vidal~\cite{vidal} through which the matrix
representation of the state of $n$ qubits ($|\Psi\ran$) is
obtained by a sequence of $n-1$ Schmidt decompositions. These
Schmidt decompositions relate to a certain ordering of the qubits,
where the $\kth$ decomposition corresponds to a partition of the
system into the first $k$ qubits and the remaining $n-k$ qubits.
Let us first write the state of the system in the computational
basis
\begin{equation}
 |\Psi\ran = \sum_{i_{1},\cdots,i_{n}=0}^{1}
 C_{i_{1}\cdots i_{n}}\,|i_{1}\ran\cdots|i_{n}\ran.
\end{equation}
The key point of the construction is the representation of the
coefficients $C_{i_{1}\cdots i_{n}}$ as a product of $n$ tensors
($\Gamma^{[k]}$) and $n-1$ vectors ($\lambda^{[k]}$)
\begin{equation}
 C_{i_{1}\cdots i_{n}} = \sum_{\alpha_{1},\cdots \alpha_{n}}
 \Gamma_{\alpha_{1}}^{[1]i_{1}} \lambda_{\alpha_{1}}^{[1]}
 \Gamma_{\alpha_{1}\alpha_{2}}^{[2]i_{2}} \lambda_{\alpha_{2}}^{[2]}
 \Gamma_{\alpha_{2}\alpha_{3}}^{[3]i_{3}} \cdots
 \Gamma_{\alpha_{n-1}}^{[n]i_{n}}
 \label{1}.
\end{equation}
Each index $\alpha_{k}$ goes from $1$ to the number of terms in
the $\kth$ Schmidt decomposition (the Schmidt number), and the
elements of the vector $\lambda^{[k]}$ are the corresponding
Schmidt coefficients. Note that the Schmidt number (or the $\log$
of this number) relating to a partition of the system can be seen
as a measure of the entanglement between the two
parts~\cite{vidal2}, and as such cannot increase under local
operations and classical communication. We denote the maximal
Schmidt number over all $n-1$ Schmidt decompositions by $\chi$.

Let us now discuss in brief Vidal's construction process. One
starts by expressing $|\Psi\ran$ using the first Schmidt
decomposition (corresponding to the partition into qubit $1$ and
the rest).
\begin{equation}
 |\Psi\ran = \sum_{\alpha_{1}} \lambda_{\alpha_{1}}^{[1]}
 |\Phi_{\alpha_{1}}^{[1]}\ran|\Phi_{\alpha_{1}}^{[2\cdots n]}\ran.
\end{equation}
Expressing the Schmidt vector $|\Phi_{\alpha_{1}}^{[1]}\ran$ in
the computational basis we obtain
\begin{equation}
 |\Psi\ran = \sum_{i_{1}\alpha_{1}} \Gamma_{\alpha_{1}}^{[1]i_{1}}
 \lambda_{\alpha_{1}}^{[1]} |i_{1}\ran|\Phi_{\alpha_{1}}^{[2\cdots n]}\ran
 \label{4}.
\end{equation}
In the next step the states $|\Phi_{\alpha_{1}}^{[2\cdots n]}\ran$
are expressed in terms of the computational basis states of the
second qubit and of  the Schmidt vectors
$|\Phi_{\alpha_{2}}^{[3\cdots n]}\ran$ corresponding to the second
Schmidt decomposition, to obtain
\begin{equation}
 |\Psi\ran = \sum_{i_{1}\alpha_{1}} \Gamma_{\alpha_{1}}^{[1]i_{1}}
 \lambda_{\alpha_{1}}^{[1]}\Gamma_{\alpha_{1}\alpha_{2}}^{[2]i_{2}}
 \lambda_{\alpha_{2}}^{[2]}|i_{1}\ran|i_{2}\ran
 |\Phi_{\alpha_{1}}^{[3\cdots n]}\ran.
 \label{6}
\end{equation}
This process can be repeated qubit by qubit until the
representation in~(\ref{1}) is obtained. The crucial point here is
that one can always express the Schmidt vectors
$|\Phi_{\alpha_{k-1}}^{[k\cdots n]}\ran$ of qubits $[k,\ldots,n]$,
obtained by the $(k-1)^{th}$ Schmidt decomposition, in terms of
the Schmidt vectors $|\Phi_{\alpha_{k}}^{[k+1 \cdots n]}\ran$,
obtained by the $k^{th}$ Schmidt decomposition. Hence we can
always write
\begin{equation}
 |\Phi_{\alpha_{k-1}}^{[k\cdots n]}\ran = \sum_{i_{k}\alpha_{k}}
 \Gamma_{\alpha_{k-1}\alpha_{k}}^{[k]i_{k}}\lambda_{\alpha_{k}}
 ^{[k]}|i_{k}\ran |\Phi_{\alpha_{k}}^{[k+1 \cdots n]}\ran.
 \label{5}
\end{equation}
It is fairly easy to see that if this was not the case and one of
$|\Phi_{\alpha_{k-1}}^{[k\cdots n]}\ran$ had a component outside
the subspace spanned by the states $
|i_k\ran|\Phi_{\alpha_{k}}^{[k+1 \cdots n]}\ran $ then the overall
state $|\Psi\ran$ could not lie within the subspace spanned by  $
|\Phi_{\alpha_{k}}^{[1\cdots k]}\ran| \Phi_{\alpha_{k}}^{[k+1
\cdots n]}\ran$, in contradiction to the $k^{th}$ Schmidt
decomposition.

A description of the state of our system in terms of the
$\Gamma^{[k]}$'s and $\lambda^{[k]}$'s would require approximately
$(2\chi^{2}+\chi)n$ parameters, instead of the $2^{n}$
coefficients ($C_{i_{1}\cdots i_{n}}$) required to represent the
state in the computational basis. The parameter which determines
the size of the description of the system is therefore $\chi$. In
general $\chi$ is of order $2^{n}$ and the MPS representation is
not very useful. However, if the state does not carry much
entanglement then $\chi$ may be smaller and the MPS representation
may be advantageous. In particular if $\chi$ scales like
poly$(n)$, than one would have an efficient description of the
state involving only poly$(n)$ parameters.

Since cluster-state computation involves only single-qubit
operations, it is easy to see that the Schmidt number associated
with any partition cannot increase, and hence that $\chi$ will not
increase during the computation. Therefore, if the initial
cluster-state has an efficient MPS representation all later states
can also be represented efficiently. Furthermore, we show that
these later representations can be obtained efficiently, and hence
prove the following: Any computation consisting of (projective)
single-qubit measurements and feed-forward on a system of $n$
qubits, where the maximal Schmidt number for all bipartitions of
the system along a certain ordering is $\chi$, can be classically
simulated at a cost of $\cal O$$(n^{2}\,\mbox{poly}(\chi))$ in
computational time and memory space.

In order to simulate a single-qubit measurement we need to
calculate the probabilities for the two outcomes, sample from the
probability distribution, project the state of the system
accordingly and then renormalize the projected state. An arbitrary
single-qubit measurement can be implemented by first applying a
single-qubit unitary, and then measuring the qubit in the
computational basis. Let us examine the effect of a single-qubit
unitary acting on the $k^{\rm th}$ qubit on the MPS representation
of the state. Clearly each of the computational basis states would
undergo the following transformation
\begin{equation}
 |i_{k}\ran
 \longrightarrow \sum_{i_{k}^{\prime}}U_{i_{k}^{\prime}i_{k}}
 |i_{k}^{\prime}\ran.
\end{equation}
This operation can be incorporated into the tensors corresponding
to the k qubit -- $\Gamma_{\alpha_{k}\alpha_{k+1}}^{[k]i_{k}}$ --
by replacing them with
\begin{equation}
 \tilde{\Gamma}_{\alpha_{k}\alpha_{k+1}}^{[k]i_{k}} =
 U_{0i_{k}}\Gamma_{\alpha_{k}\alpha_{k+1}}^{[k]0} +
 U_{1i_{k}}\Gamma_{\alpha_{k}\alpha_{k+1}}^{[k]1}.
\end{equation}
Updating the MPS representation after a single-qubit unitary would
therefore take $\cal O$$(\chi^{2})$ basic operations.

The probabilities of the outcomes in the computational basis can
be easily calculated from the following representation of the
state of the system, obtained by using the $(k-1)^{\rm th}$
Schmidt decomposition, and equation (\ref{5}),
\begin{equation}
 \sum_{i_{k}} \sum_{\alpha_{k-1}\alpha_{k}}
 \lambda_{\alpha_{k-1}}^{[k-1]}\Gamma_{\alpha_{k-1}\alpha_{k}}^{[k]i_{k}}
 \lambda_{\alpha_{k}}^{[k]}
 |\Phi^{[1\cdots k-1]}_{\alpha_{l-1}}\ran |i_{k}\ran
 |\Phi^{[k+1\cdots n]}_{\alpha_{l}}\ran.
\end{equation}
The probability $p(i_{k})$ for receiving the outcome $i_{k}$ in a
measurement of the $k$ qubit is therefore obtained from the tensor
$\Gamma_{\alpha_{k-1}\alpha_{k}}^{[k]i_{k}}$ and the vectors
$\lambda_{\alpha_{k-1}}^{[k-1]}$ and $\lambda_{\alpha_{k}}^{[k]}$.
Defining
\begin{equation}
 A_{\alpha_{k-1}\alpha_{k}}^{i_k} = \lambda_{\alpha_{k-1}}^{[k-1]}
 \Gamma_{\alpha_{k-1}\alpha_{k}}^{[k]i_{k}}\lambda_{\alpha_{k}}^{[k]}
\end{equation}
and using the orthogonality of the Schmidt vectors, we have
\begin{equation}
 p(i_{k}) = \sum_{\alpha_{k-1}\alpha_{k}}
 |A_{\alpha_{k-1}\alpha_{k}}^{i_k}|^{2}.
\end{equation}

Sampling from this probability distribution, and receiving outcome
$|r_{k}\ran$, the state of the system after projection and
renormalisation will be $|r_k\ran |\Psi^{\prime}\ran$, where
\begin{equation}
 |\Psi^{\prime}\ran = \frac{1}{\sqrt{p(r_{k})}}\sum_{\alpha_{k-1}\alpha_{k}}
 A_{\alpha_{k-1}\alpha_{k}}^{r_{k}}
 |\Phi^{[1\cdots k-1]}_{\alpha_{k-1}}\ran
 |\Phi^{[k+1\cdots n]}_{\alpha_{k}}\ran.
 \label{12}
\end{equation}
In what follows, we leave out the state of the measured qubits
(which remain in a product state with the rest of the system), and
consider the MPS representation of the remaining qubits. This
representation must now be updated since the $\lambda$'s and
$\Gamma$'s above do not correspond to Schmidt decompositions of
$|\Psi^{\prime}\ran$, and in order to be able to calculate the
probability distribution for the measurement of the remaining
qubits efficiently, the correct MPS representation must be
recovered. Indeed as we consider a general $n$-qubit state (where
the measurement of one qubit might affect all other qubits) all of
the $\Gamma$'s and $\lambda$'s must be updated. Tracing over
qubits $1,\ldots,k-1$ we obtain the $\chi$ by $\chi$ reduced
density matrix $\rho^{[k+1\cdots n]}$
\begin{equation}
 \rho_{\alpha_{k}\alpha_{k}^{\prime}}^{[k+1\cdots n]} =
 \sum_{\alpha_{k-1}} A_{\alpha_{k-1}\alpha_{k}}^{r_k}
 (A_{\alpha_{k-1}\alpha_{k}^{\prime}}^{r_k} )^{*}.
\end{equation}
Given the reduced density matrix we can its find eigenvalues
$\tlambda_{\beta_{k-1}}^{[k-1]}$, which are the Schmidt
coefficients for the above partition. We can also find its
eigenvectors $M_{\beta_{k-1}\alpha_{k}}$, which represent the new
Schmidt vectors $|\Phi_{\beta_{k-1}}^{[k+1\cdots n]}\ran$ in the
basis of the old Schmidt vectors:
\begin{equation}
 |\Phi_{\beta_{k-1}}^{[k+1\cdots n]}\ran =
 \sum_{\alpha_{k}} M_{\beta_{k-1}\alpha_{k}}
 |\Phi_{\alpha_{k}}^{[k+1\cdots n]}\ran.
 \label{13}
\end{equation}
To calculate the Schmidt coefficients
($\tlambda_{\beta_{k}}^{[k]}$) for the next partition, between
qubits $[1,\ldots, k-1, k+1]$ and $[k+1,\ldots, n]$, we write the
state of system as follows
\begin{eqnarray}
 |\Psi^{\prime}\ran &=& \sum_{\beta_{k-1}}\lambda_{\beta_{k-1}}^{[k-1]}
 |\Phi_{\beta_{k-1}}^{[1\cdots k-1]}\ran
 |\Phi_{\beta_{k-1}}^{[k+1\cdots n]}\ran \nonumber \\
 &=& \sum_{\beta_{k-1} \alpha_{k}} \lambda_{\beta_{k-1}}^{[k-1]}
  M_{\beta_{k-1}\alpha_{k}}|\Phi_{\beta_{k-1}}^{[1\cdots k-1]}\ran
 |\Phi_{\alpha_{k}}^{[k+1\cdots n]}\ran  \label{expand_psi_eqn} \\
 &=& \sum_{\beta_{k-1} \alpha_{k+1} \atop i_{k+1}}
 B_{\beta_{k-1} \alpha_{k+1}}^{i_{k+1}}
 |\Phi_{\beta_{k-1}}^{[1\cdots k-1]}\ran |i_{k+1}\ran
 |\Phi_{\alpha_{k+1}}^{[k+2\cdots n]}\ran \nonumber
\end{eqnarray}
where
\begin{equation}
 B_{\beta_{k-1}\alpha_{k+1}}^{i_{k+1}} = \sum_{\alpha_{k}}
 \lambda_{\beta_{k-1}}^{[k-1]} M_{\beta_{k-1}\alpha_{k}}
 \Gamma_{\alpha_{k}\alpha_{k+1}}^{[k+1]i_{k+1}}
 \lambda_{\alpha_{k+1}}^{[k+1]}.
\end{equation}
We can now write the reduced density matrix of qubits $[k+2,\ldots
n]$ in terms of the tensor $B$
\begin{equation}
 \rho_{\alpha_{k+1}\alpha_{k+1}^{\prime}}^{[k+2\cdots n]} =
 \sum_{\beta_{k-1} i_{k+1}} B_{\beta_{k-1}\alpha_{k+1}}^{i_{k+1}}
 (B_{\beta_{k-1}\alpha_{k+1}^{\prime}}^{i_{k+1}})^{*}.
\end{equation}
Having calculated $\rho^{[k+2\cdots n]}$ we can now find its
eigenvalues $\tlambda_{\beta_{k}}^{[k]}$ (the Schmidt
coefficients) and its eigenvectors $M_{\beta_{k}\alpha_{k+1}}$,
which represent the new Schmidt vectors
$|\Phi_{\beta_{k}}^{[k+2\cdots n]}\ran$ in the basis of the old
Schmidt vectors $|\Phi_{\alpha_{k+1}}^{[k+2\cdots n]}\ran$, as in
(\ref{13}).

The relation between $|\Phi_{\beta_{k}}^{[k+2\cdots n]}\ran$ and
$|\Phi_{\beta_{k-1}}^{[k+1\cdots n]}\ran$ defines the new tensors
$\tGamma_{\beta_{k-1}\beta_{k}}^{[k+1]i_{k+1}}$ corresponding to
qubit $k+1$. Examining (\ref{5}) we see that
\begin{eqnarray}
\tGamma_{\beta_{k-1} \beta_{k}}^{[k+1]i_{k+1}} &=&
\frac{1}{\tlambda_{\beta{k}}^{[k]}} \langle
i_{k+1},\Phi_{\beta_{k}}^{[k+2\cdots n]} \mid
 \Phi_{\beta_{k-1}}^{[k+1\cdots n]} \ran \label{tGamma_eqn} \\ &=&
 \frac{1}{\tlambda_{\beta_{k}}^{[k]}} \sum_{\alpha_{k}\alpha_{k+1}}
 M_{\beta_{k}\alpha_{k+1}}^{*} M_{\beta_{k-1}\alpha_{k}}
 \Gamma_{\alpha_{k} \alpha_{k+1}}^{[k+1]i_{k+1}}
 \lambda_{\alpha_{k+1}}^{[k+1]} \nonumber
\end{eqnarray}
where we have used (\ref{13}) and the corresponding relation for
$|\Phi_{\beta_{k}}^{[k+2\cdots n]}\ran$ in the last line.

We can proceed in the same manner to obtain all the reduced
density matrices in one direction ($\rho^{[k+3\cdots n]}$ to
$\rho^{[n]}$) and their eigenvalues and eigenvectors. At each step
the reduced density matrix is given as a function of the
eigenvectors and  eigenvalues obtained in the previous step (as
well as the old $\Gamma$'s and $\lambda$'s). The new $\tGamma$'s
are found using the eigenvectors of two consecutive steps as in
(\ref{tGamma_eqn}). In order to update the MPS representation in
the other direction we first express
$|\Phi_{\beta_{k-1}}^{[1\cdots k-1]}\ran$ in terms of
$|\Phi_{\alpha_{k-1}}^{[1\cdots k-1]}\ran$ using (\ref{12}),
(\ref{13}) and
\begin{equation}
|\Phi_{\beta_{k-1}}^{[1\cdots k-1]}\ran =
\frac{1}{\tlambda_{\beta_{k-1}}} \langle
\Phi_{\beta_{k-1}}^{[k+1\cdots n]} |\Psi' \ran.\label{left1_eqn}
\end{equation}
then re-express these states in terms of
$|\Phi_{\alpha_{k-2}}^{[1\cdots k-2]} \ran|i_{k-1} \ran$ using
\begin{equation}
|\Phi_{\alpha_{k-1}}^{[1\cdots k-1]}\ran =
\sum_{i_{k-1}\alpha_{k-2}}\lambda_{\alpha_{k-2}}^{[k-2]}
\Gamma_{\alpha_{k-2}\alpha_{k-1}}^{[k-1]i_{k-1}}
|\Phi_{\alpha_{k-2}}^{[1 \cdots k-2]}\ran |i_{k-1}\ran.
\label{left2_eqn}
\end{equation}
Expanding $|\Psi'\ran$ as in (\ref{expand_psi_eqn}) using
(\ref{left1_eqn}) and (\ref{left2_eqn}), and tracing over qubits
$[k-1,k+1,\ldots,n]$ we obtain $\rho^{[1\cdots k-2]}$. The
eigenvalues and eigenvectors of this matrix can be used to obtain
$\tlambda_{\beta_{k-2}}^{[k-2]}$ and
$\tGamma^{[k-1]i_{k-1}}_{\beta_{k-2} \beta_{k-1}}$ in a similar
way to before, and repeating this procedure we can obtain all of
the remaining $\tGamma$'s and $\tlambda$'s.

In each step of the updating procedure we deal with a constant
number of (at most) $\chi$ by $\chi$ matrices, requiring
poly$(\chi)$ basic operations, and hence updating the whole state
after a measurement requires $n\,$poly$(\chi)$ basic operations.
As we can apply at most $n$ single-qubit measurements to the
state, the overall computation can be simulated with $\cal
O$$(n^{2}\,\mbox{poly}(\chi))$ computational resources. Note that
we do not include the cost of computing the feed-forward from the
measurement results, as this is common to both the quantum
computation and the classical simulation (and for standard
cluster-state computation can be computed efficiently).

In the above procedure we simulate the single-qubit measurements
in the same order as they are measured in the actual computation.
This is usually independent of the order in which qubits in the
MPS representation are numbered, which we choose so as to minimize
$\chi$. However, if we can number the MPS representation in the
same order as the qubits are measured without significantly
increasing $\chi$, then the simulation can be considerably
simplified. In this case we do not have to update the MPS
representation after the measurements, as the probability
distribution can be calculated directly from the projected state.
The cost of the simulation in this case would be $\cal
O$$(n\,\chi^{2})$.

Let us now consider a cluster state in the shape of a rectangular
grid of width $d$ and length $l>d$ where each qubit is entangled
to all its nearest neighbors. In order to construct our MPS
representation, we choose the following ordering of the qubits: We
start from the qubit in the top-left corner and number the qubits
column by column until we reach the qubit in the bottom-right
corner. In this case, the maximal Schmidt number $\chi$ for all
bipartitions of the system is $2^{d}$. To prove this, we consider
a typical bipartition, and an incomplete cluster, similar to the
original but where the $2d$ qubits next to the partition are only
entangled with each other (and not with the rest of the cluster),
as shown in figure 1. The original cluster can be constructed from
the incomplete cluster by local unitary operations (that is,
operations which act on only one side of the partition) and
therefore both have the same Schmidt number $S$ with respect to
this partition. The state of the incomplete cluster can be written
as $|\Psi_{A}\ran|P\ran|\Psi_{B}\ran$, where $|P\ran$ stands for
the entangled state of the $2d$ qubits next to the partition and
$|\Psi_{A/B}\ran$ are the states of the remaining qubits on the
left and right sides respectively. Since, $|P\ran$ contains only
$2d$ qubits, it is clear that the $S \leq 2^{d}$. In fact one can
easily verify that $S = 2^{d}$, since $|P\ran$ itself can be
constructed from $d$ maximally entangled pairs of qubits by local
unitary operations. As no bipartition has a greater Schmidt number
than $S$, we have $\chi = 2^{d}$.

\begin{figure}\begin{center}
\epsfig{file=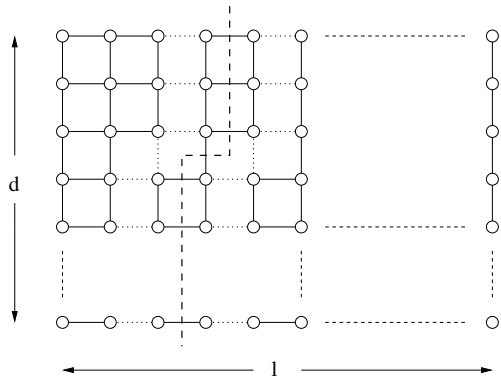} \caption{\small  A rectangular cluster
state. The solid lines represent entanglement (CPHASE
connections), the thick dashed line shows a typical partition of
the cluster, and the dotted lines show the connections that do not
exit in the incomplete cluster.}
\end{center}\end{figure}

In general we expect both $d$ and $l$ to scale as poly$(N)$ where
$N$ is the number of logical qubits, and hence the above
simulation would require exponential resources. However, any
cluster-state computation implemented by a grid of physical qubits
with limited width, that is, where $d$ scales like (at most)
$\log(N)$, can be simulated at a polynomial cost in time and
memory space.

We can also extend our approach to more general cluster-states, in
which non-neighbouring qubits are connected by CPHASE operations,
as long as these connections have limited range. Consider a
cluster where the vertical distance  between connected qubits
(across the width $d$) is not limited, and the horizontal distance
is limited by $r$. That is, a qubit in column $k$ may be connected
to any qubit in columns $\{k-r,\ldots,k+r\}$. Looking at a typical
partition where the left side consists of the first $k-1$ columns
and part of the $\kth$ column, and the right side contains the
rest of the cluster, we consider an incomplete cluster where the
block consisting of columns $\{k-r,\ldots,k+r\}$ is isolated from
the rest of the cluster (i.e. in the incomplete cluster the qubits
of this block have the same connections between themselves but no
connections to the rest of the cluster). As in the previous case,
the state of the incomplete cluster is given by
$|\Psi_{A}\ran|P^{\prime}\ran|\Psi_{B}\ran$ where
$|P^{\prime}\ran$ is the state of the isolated block.
$|P^{\prime}\ran$ consists of only $(2r+1)d$ qubits, so the
Schmidt number for the partition must be less than $2^{(r+1/2)d}$.
As before, the original cluster can be recovered by local CPHASE
operations, hence $\chi \le 2^{(r+1/2)d}$. It is therefore clear
that any cluster-state computation on a rectangular grid can be
efficiently simulated with a classical computer as long as either
$d$ (the width of the cluster) or $r$ (the range of the
connections) scales like $\log(N)$, and the other is constant.

In our general procedure we did not specify to which qubits the
input is introduced and in which order the measurements are
performed in the computation. Given an efficient MPS
representation of the initial state of the cluster any computation
can be simulated efficiently. Thus, considering the $l\times d$
rectangular grid above, we can allocate a column of physical
qubits for each logical qubit (measuring the physical qubits, say,
from top to bottom). Simulating a quantum gate array in this way,
$l$ would be proportional the number of logical qubits $N$, $d$
would be proportional to the depth of the computation (i.e. the
number of time-steps), and $r$ would be proportional to the range
over which gates can act. Therefore, we can also state that any
quantum computation in the gate array model where either the depth
of the computation or the range of the interaction scales like
$\log(N)$, while the other is bounded by a constant, can be
efficiently simulated on a classical computer.

\acknowledgments     

The authors wish to thank Sandu Popescu, Richard Josza, Noah
Linden and Emma Podnieks for fruitful discussions. The work of
N.~Y. was supported by UK EPSRC grant (GR/527405/01), and A.J.S.
was supported by the UK EPSRC's ``QIP IRC'' project.

\end{document}